# Molecular Dynamic Study of Local Interfacial Thermal Resistance of Solid-Liquid and Solid-Solid Interfaces: Water and Nanotextured Surface

*Yoshitaka Ueki\*, Satoshi Matsuo, Masahiko Shibahara*

Department of Mechanical Engineering, Osaka University, 2-1 Yamadaoka, Suita, Osaka 565-0871, Japan

ABSTRACT: Degradation in the performances of air conditioners and refrigerators is caused by frost formation and adhesion on the surface. In the present study, by means of the classical molecular dynamics simulation, we investigate how and how much the nanotextured surface characteristics, such as surface wettability and geometry, influenced the interfacial thermal resistance (ITR) between the solid wall and the water/ice. The ITR of the interfacial region was comparable in both the water and the ice states. As the nanostructure gaps became narrower, the ITR of the interfacial region decreased. The local ITR had a weak negative correlation with the local $H_2O$ molecule density regardless of the phase of the $H_2O$ molecules. The local ITR decreased as the local density increased. A greater amount of the thermal energy was transferred through the material interface by means of the intermolecular interaction when more $H_2O$ molecules were located in the proximity area, which was closer to the Pt solid wall than the first adsorption layer peak. When the $H_2O$ molecules were in the crystal form on the solid wall, the proximity molecules decreased, and then the local ITR significantly increased.

## 1. INTRODUCTION

Icephobicity of heat transfer surfaces is able to mitigate degradation in performances of air conditioners and refrigerators caused by frost formation and adhesion on the surface. Defrosting, which heats up the heat transfer surfaces and melt the frost, is necessary be done when the frost forms on the heat transfer surfaces. One possible approach to meet the growing demand for higher cooling performance and efficiency is to suppress the frost formation and ice adhesion to solid surfaces. Superhydrophobicity has been considered one of the candidates to gain the icephobicity of the heat transfer surfaces. Superhydrophobic surfaces can enhance to slip off condensed droplets before freezing. Heterogeneous surface geometry and chemistry had been employed to achieve efficient dropwise condensation (e.g., [1]). Some experimental studies (e.g. [2]) had demonstrated that superhydrophobic nanostructured surfaces induced condensed droplets coalesced and jumped off from the heat transfer surfaces, which significantly enhanced the condensation heat transfer. Other studies (e.g. [3]) had shown that slippery lubricant-infused porous surfaces (SLIPSs) could achieve high condensation heat transfer. The SLIPSs can promote the condensed droplets slip off from the heat transfer surfaces. Although the droplet-repelling surfaces work effectively to achieve high condensation heat transfer performances, they cannot gain the perfect icephobicity. It is because just superhydrophobic surfaces cannot always become ice-repellent, and therefore their use as anti-ice materials may be limited. Previous studies showed that the ice formation formed even in gaps of the micro-textured superhydrophobic surface, and then an ice adhesion to the solid surface increased to become stronger than that of flat surfaces [4], or almost similar to it in terms of the adhesion strength [5]. Nanotextured superhydrophobic surfaces [6] could significantly reduce the ice adhesion strength further down below that of the smooth hydrophobic surfaces. Still, ice adhesion is unavoidable because ice nucleation happens. It means that defrosting is the required procedure to regenerate the heat transfer performance of the air conditioners and the refrigerators. As described above, the condensed-water repellency alone is not sufficient for the icephobicity. It requires the ability to significantly suppress the ice nucleation, impede the frost formation, and reduce the ice adhesion forces. As of now, there have been no perfect solutions to gain passive icephobicity, especially in the natural conditions where ice accretion occurs over a wide range of temperatures, humidity levels, and wind conditions [7].

Besides the heat transfer equipment, aircraft require to manage an adhesion and buildup of the frost, ice, snow, and slush on themselves since freezing the aircraft wings degrades aircraft lift force and flight stability. In severe winter seasons, before aircraft departures, deicing and anti-icing fluids are sprayed on parts of the aircraft, including the wings, tail, antennas, and sensors. During the flights, the aircraft need to keep the ice from accumulating on their surfaces. On some commercial aircraft, the hot air generated by engines is transferred to the wings and the tails to maintain them warm and prevent icing. On others, electric heaters are equipped with wings to prevent icing.

In summary, defrosting and deicing are commonly the important and inevitable heat transfer process in the above-mentioned machinery. Based on this background, we focus on interfacial thermal resistance (ITR) between water and the nanotextured surfaces. It is because, especially in the case of the nanotextured surfaces, the ITR is one of the important factors to determine the processing time of the defrosting and the deicing

In nanoscale systems, the ITR plays an important role in thermal energy transports through solid-liquid interfaces and the solid-solid interfaces. The ITR changes with its interface geometry, surface wettability (or intermolecular interaction between each phase component), and other interface conditions. For example, in the case of the solid-liquid interface, the ITR of the hydrophilic and hydrophobic interface, which were functionalized with a self-assembled monolayer (SAM), was experimentally evaluated by means of time-domain thermoreflectance [8]. It showed that the ITR at hydrophobic interfaces was 2-3 times greater than that at hydrophilic interfaces. Nanotextured interface geometry affects the solid-liquid ITR. Molecular dynamics (MD) studies (e.g. [9]) employed nanoslit structure on the solid-liquid interface and showed that the nanoslit was able to reduce the ITR depending on the nanostructure geometry. Other surface modifications are also able to change the ITR. For example, previous MD studies showed nanoscale surface roughness [10], nanoparticle layers [11], graphene sheets [12], and SAMs [13] on the solid-liquid interface influenced the ITR. Some spectral analyses [14-16] had been conducted and showed that phonon density of states (VOS) and spectral heat flux could give further insight into the thermal transport at the solid-liquid interfaces.

However, we have a limited understanding of the ITR between the ice and the heat transfer surface. Therefore, in the present study, we focused on the ITR between water (at the liquid and solid phase) and the solid wall of the heat transfer surface. Furthermore, we investigated the influences of surface wettability and geometry on the local thermal energy transfer at the material interfaces in the case of the nanotextured heat transfer surface by means of the classical non-equilibrium molecular dynamics simulations.

## 2. NUMERICAL METHOD

**Figure 1** illustrates an example of the simulation system employed in the present study. The unit cell was 7.99 nm in *x*- and *y*- directions. The size of the unit cell in z-direction changed in the range of from 9.50 to 10.6 nm, in order to maintain the system pressure on average to be in the range of from -3 to 5 MPa. The periodic boundary condition was applied in *x*- and *y*- directions. The nanoslit dimension was $2.00 \times 7.99 \times 2.00$ nm$^3$. We employed three types of surface geometry of the solid wall, as shown in **Figure 2**. In the case of Nano1 (see **Figure 2** (b)), an opening of the nanoslit was 5.99 nm. Moreover, in the case of Nano2 (see **Figure 2** (c)), that of the nanoslit was 2.00 nm. The solid wall consisted of Pt atoms. The fluid molecules were H$_2$O. The intermolecular potential for the H$_2$O molecules was mW potential [17], which was expressed as follow:

$$E = \sum_i \sum_{j>i} \varphi_2(r_{ij}) + \sum_i \sum_{j \neq i} \sum_{k>j} \varphi_3(r_{ij}, r_{ik}, \theta_{ijk}) \tag{1}$$

$$\varphi_2(r_{ij}) = A\varepsilon \left[ B\left(\frac{\sigma}{r_{ij}}\right)^p - \left(\frac{\sigma}{r_{ij}}\right)^q \right] \exp\left(\frac{\sigma}{r_{ij} - a\sigma}\right) \tag{2}$$

$$\varphi_3(r_{ij}, r_{ik}, \theta_{ijk}) = \lambda\varepsilon \left[\cos\theta_{ijk} - \cos\theta_0\right]^2 \exp\left(\frac{\gamma\sigma}{r_{ij} - a\sigma}\right) \exp\left(\frac{\gamma\sigma}{r_{ik} - a\sigma}\right) \tag{3}$$

The mW model is a coarse-grained H$_2$O model and has been employed to investigate the H$_2$O phase change phenomena and characteristics [18-20]. For the pairs of Pt atoms, we employed 12-6 Lennard-Jones intermolecular potential as expressed by Eq. (4). Moreover, for the pairs of the Pt atom and the H$_2$O molecule, we employed the 12-6 Lennard-Jones potential multiplied with the interaction parameter *α*, as expressed by Eq. (5).

$$\phi_{ij}\left(r_{ij}\right) = 4\varepsilon\left\{\left(\frac{\sigma}{r_{ij}}\right)^{12} - \left(\frac{\sigma}{r_{ij}}\right)^{6}\right\} \quad (4)$$

$$\phi_{ij}\left(r_{ij}\right) = 4\alpha\varepsilon\left\{\left(\frac{\sigma}{r_{ij}}\right)^{12} - \left(\frac{\sigma}{r_{ij}}\right)^{6}\right\} \quad (5)$$

A higher value of $\alpha$ corresponded to better wettability. In the case of $\alpha = 0.02$, the contact angle of the H$_2$O droplet on the Pt flat surface was 130º. In the case of $\alpha = 0.05$, the contact angle was 75º. The Lorentz-Berthelot combining rule was employed to determine the potential parameters of the pairs of Pt and H$_2$O. The mW potential parameters [17] were shown in Table 1. The Lennard-Jones parameters employed in the present study [21] were shown in Table 2. The mass of the Pt atoms was 195 g/mol, and that of the H$_2$O molecules was 18.0 g/mol. The time step was 1.0 fs. The cut-off distance was 1.8 $\sigma$ for the pair of the mW potential and 3.0 $\sigma$ for the pair of the Lennard-Jones (LJ) potential. The leap-frog method was employed for the numerical integration. The numbers of the particles employed in the simulation system were shown in Table 3.

**Table 1. mW potential parameters** [17].

| $\varepsilon$ kcal/mol | $\sigma$ nm | $a$ | $\lambda$ | $\gamma$ |
|---|---|---|---|---|
| 6.189 | 0.23925 | 1.80 | 23.15 | 1.20 |
| $\cos\theta_0$ | $A$ | $B$ | $p$ | $q$ |
| −0.33333 | 7.04956 | 0.602225 | 4.0 | 0.0 |

**Table 2. Lennard-Jones potential parameters** [21].

|  | $\sigma$ nm | $\varepsilon$ J |
|---|---|---|
| Pt - Pt | 0.254 | $1.09 \times 10^{-19}$ |
| H$_2$O - Pt | 0.247 | $6.82 \times 10^{-20}$ |

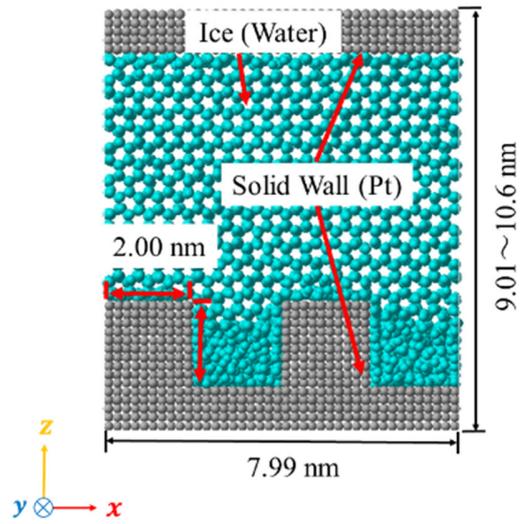

**Figure 1. Simulation model.**

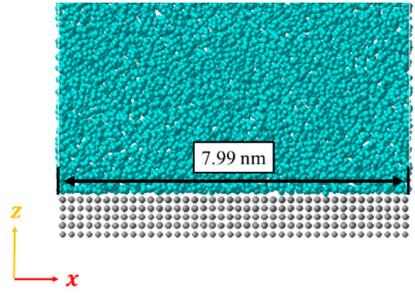

**(a) Flat.**

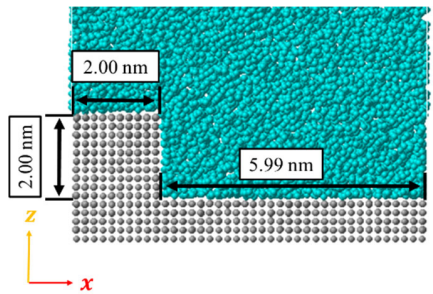

**(b) Nano1.**

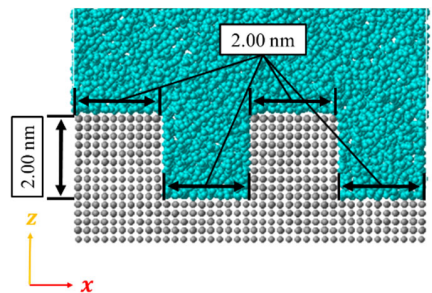

**(c) Nano2.**

**Figure 2. Surface geometry.**

Table 3. Numbers of particles in simulation system.

|  |  |  | H$_2$O | Pt |
|---|---|---|---|---|
| Water | $\alpha = 0.02$ | Flat | 16191 | 8000 |
|  |  | Nano1 | 15125 | 10000 |
|  |  | Nano2 | 13961 | 12000 |
|  | $\alpha = 0.05$ | Flat | 15125 | 8000 |
|  |  | Nano1 | 15125 | 10000 |
|  |  | Nano2 | 13961 | 12000 |
| Ice | $\alpha = 0.02$ | Flat | 16191 | 8000 |
|  |  | Nano1 | 15125 | 10000 |
|  |  | Nano2 | 13961 | 12000 |
|  | $\alpha = 0.05$ | Flat | 16191 | 8000 |
|  |  | Nano1 | 15125 | 10000 |
|  |  | Nano2 | 13961 | 12000 |

The simulation procedures in the present study were as follows. We employed the Langevin method for thermostatting the Pt walls and the velocity scaling for the H$_2$O molecules. In the case of the water state, in the initial temperature control time of 0.1 ns, the upper Pt wall was maintained at 350 K, and the lower wall was maintained at 300 K. The fluid molecules were controlled to be at 325 K. Next, the velocity scaling of the H$_2$O molecules was off. And then, a barostatting procedure was done by adjusting the location of the solid upper wall in $z$-direction so that the pressure applied on the upper wall was in the range of from -3 to 5 MPa. After this, the thermostatting of the upper and lower walls continued for 2.0 ns so that the system reached a steady state. In the next 5.0 ns, the main calculation and data acquisition were performed. In the case of the ice state, the thermostatting temperatures were different. The upper Pt wall was maintained at 270 K, and the lower wall was maintained at 220 K. The fluid molecules were controlled to be at 245 K. The other procedures were identical to the case of the water state.

We defined the interfacial region, which contains the nanotexture and the H$_2$O molecules with the thickness of the nanotexture height. Based on this definition, the ITR was evaluated in the previous study (e.g. [9]). The ITR of the interfacial region was calculated as follows:

$$R_{\text{th, region}} = \frac{\Delta T_{\text{region}}}{q} \qquad (6)$$

$R_{\text{th, region}}$ is the effective thermal resistance with the thickness of the interfacial region. $\Delta T_{\text{region}}$ was evaluated from the temperature profiles of Pt and H$_2$O, as shown in **Figure 3**. It was the temperature difference between the H$_2$O temperature at the location of the nanotexture top and the Pt one at the location of the nanotexture bottom. $q$ denotes the heat flux passing through the simulation system, and it is identical to the heat flux given and received by the Langevin layers.

In the present study, we divided regions of the solid-liquid and the solid-solid interfaces into fine interrogation volumes. Figure 4 illustrates the definitions of the interrogation volumes. At both the Pt and H$_2$O sides, interfacial interrogation volumes were prepared. In the case of the nanotextured surface, the interfacial interrogation volumes were defined by following the surface geometry. In this case, there were some of the regions where the H$_2$O molecules were not in contact with Pt, such as the bottom corner regions. Then, we were not able to define them as the interfacial regions and so excluded them from the interfacial interrogation volumes. Based on the present local region segmentation, as the phase interface, we defined the planes where the neighboring Pt and H$_2$O interrogation volumes were facing. Then, we categorized the interfaces into the following local interfaces: Bottom Corner, Top Corner, Nanost. Top, Nanost. Side and Nanost. Bottom, as shown in **Figure 4**. In the $x$- and $z$-directions, the interrogation volumes were 0.40 nm, which corresponded to approximately the lattice constant of Pt. In the *y*-direction, the interrogation volume was identical to the simulation system.

The local ITR of $R_{th, m}$ was calculated from the following equation:

$$R_{th,m} = \frac{\Delta T_m}{q_m} \quad (7)$$

$\Delta T_m$ denotes the temperature difference between the neighboring interrogation volumes over the interface. $q_m$ denotes the heat flux flowing into the wall-side interrogation volumes. It was calculated from the following equation.

$$q_m = \frac{1}{A_m} \sum_{\substack{i \in Pt \\ Pt \in m}} \sum_{j \in H_2O} \left( \mathbf{F}_{ij} \cdot \mathbf{v}_i \right) \quad (8)$$

$v_i$ denotes the velocity vector of the Pt atoms. $F_{ij}$ denotes the force vector from the H$_2$O molecules to the Pt atoms. $A_m$ denotes the interfacial surface area. m corresponds to the interfacial interrogation volumes. In the case of Top Corner, the interfacial surface areas in contact with the H$_2$O molecules were two times greater than those of others.

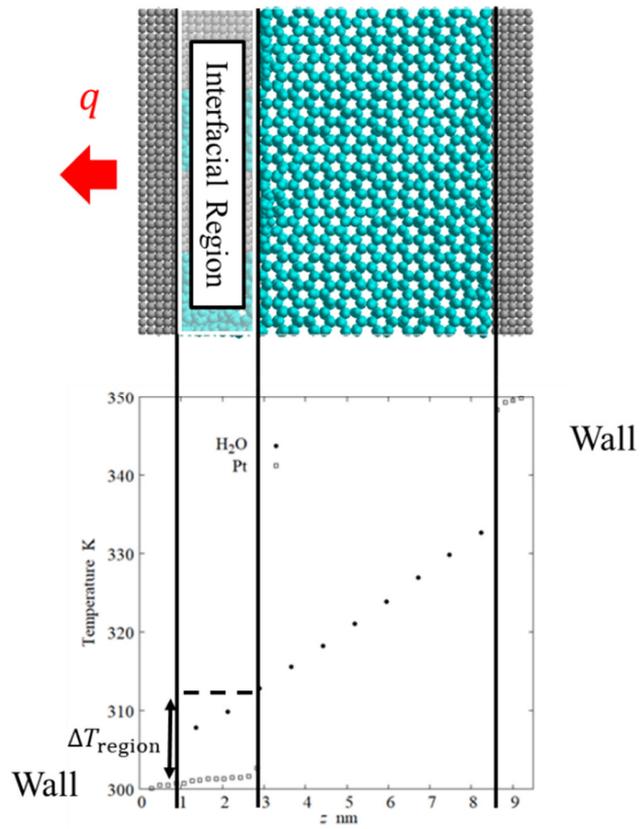

**Figure 3. Definition of interfacial region.**

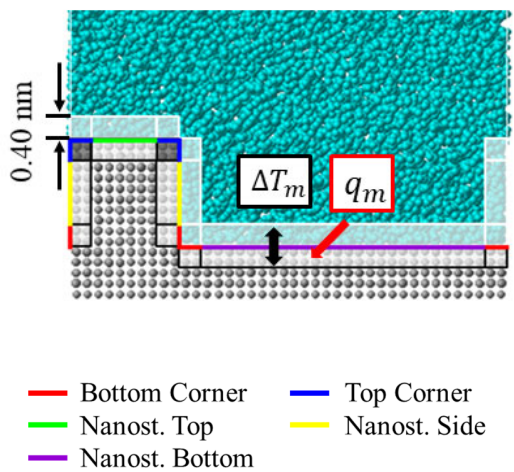

— Bottom Corner — Top Corner
— Nanost. Top — Nanost. Side
— Nanost. Bottom

**Figure 4. Definition of interfacial interrogation volumes.**

## 3. Numerical Results and Discussion

The surface wettability significantly influences the local density of the fluid molecules and the ITR. Hereafter, we described the case of $\alpha = 0.05$, which corresponded to the hydrophilic condition. **Figure 5** illustrates snapshots of the initial simulation system in the case of the water state. The Wenzel state was achieved in the case of the nanotextured surfaces. **Figure 6** illustrates snapshots of the initial simulation system in the case of the ice state. Note that the $H_2O$ molecules were located in the nanoslit openings. In **Figure 6** (a), the solid $H_2O$ structures were different on the top and bottom wall surfaces, which were facing the $H_2O$ molecules. On the top surface, the $H_2O$ molecules were in a crystal form. On the other hand, on the bottom surface, they were in an amorphous form. The difference in the structures was due to the way to produce the initial states of the system. In the present study, to produce the initial state of the system, we initiated the $H_2O$ molecules crystalization downward toward the bottom wall. It was because how to proceed with the solidification in the vicinity of the nanotextured surfaces had been known. In the case of Nano1 (see **Figure 6** (b)), near the bottom wall and the nanostructure sidewalls, the amorphous layer was formed. However, the crystal state was maintained even on the top of the wall nanostructures. In the case of Nano2 (see **Figure 6** (c)), near the top of the nanostructure, the crystal structure was disordered slightly. In between the nanostructures, the $H_2O$ molecules were entirely in the amorphous form due to the narrow opening of the nanostructures. Notably, such small nanostructure openings can change the solid-state structures of the $H_2O$ molecules.

### 3.1 Temperature Distributions

**Figures 7** and **8** illustrate two-dimensional temperature distributions in the *x-z* plane near the bottom wall. The averaging time for the temperature distributions was 5.0 ns. In the case of the water state, quasi-one-dimensional temperature distributions were formed in the *z*-direction. On the other hand, in the case of the ice state, the temperature gradient in the crystal state region became gentle. Notably, in the case of Nano2, where the ice structure was in the amorphous form, the ice temperature on the nanostructure bottom was lower than that in the case of Nano1. Figures 9 and 10 illustrate the one-dimensional temperature distributions in the *z*-direction. The wall temperature distributions consisted of the averaged temperatures of each layer of the Pt atoms except the fixed outermost layers. The $H_2O$ volumes were divided into ten interrogation volumes in the *z*-direction. From each of the interrogation volumes, the $H_2O$ temperature distributions were calculated. The temperature gradients of the $H_2O$ regions were evaluated except those neighboring to the walls and in between the nanostructures. The evaluated temperature gradients were shown in **Figures 9** and **10**. We found that the $H_2O$ temperature gradients became steeper as the nanostructure openings became narrower in both cases of the water and the ice states. It was because of an increase in the intermolecular interaction between the $H_2O$ molecules and the Pt atoms consisting of the nanostructure and the bottom wall, which cooled down the $H_2O$ molecules near the bottom wall.

From the one-dimensional temperature distributions, we could find some differences between the water and the ice states consistently with the two-dimensional temperature distributions. In the case of the water state, the quasi-one-dimensional temperature distributions were formed in the *z*-direction. In the case of the ice state, the temperature gradients changed between the nanostructures. Since the thermal conductivity of the crystal structure was higher than that of the amorphous structure, the temperature gradients of the crystal region became gentle compared to those in the amorphous region. This tendency was obvious in the case of Nano2.

## 3.2 Density Distributions

**Figures 11** and **12** show two-dimensional density distributions of the water and the ice states in the *x-z* plane. From **Figure 11**, we found that the adsorption layers of the fluid molecules were formed on the solid walls and along the nanostructures. Moreover, between the first and the second peaks of the adsorption layers, there were low-density layers in all of the cases. From **Figure 12** (a), we found that, in the ice state on the flat surface, the crystal structure was formed and vibrated in the horizontal directions with respect to the flat surface. On the other hand, in the case of the nanotextured surfaces, the horizontal vibration was suppressed. Because of this, the nanotextured surfaces contributed to the vibration suppression. From **Figure 12** (b), we found that, in the case of Nano1, the ice around the bottom corner was in the amorphous form, as seen in **Figure 6** (b). From **Figure 12** (c), in the case of Nano2, the ice in between the nanostructures was in the amorphous form, as seen in **Figure 6** (c). Comparing the density of the $H_2O$ molecules in between the nanostructures in the water and ice states (see **Figures 11** (c) and **12** (c)), it was quite different from each other. In the ice state, coarse destiny spots were found in a slightly-ordered array in the middle of the nanostructure gaps. On the other hand, in the water case, the fluid adsorption layering structure was formed in the nanostructure gaps. It indicated that the ice in the nanostructure gaps was not in the form of the liquid nor the adsorption layer but in the amorphous solid form. In addition, the present study showed that the nanotextured surface was able to transform the solidification structure near the solid walls.

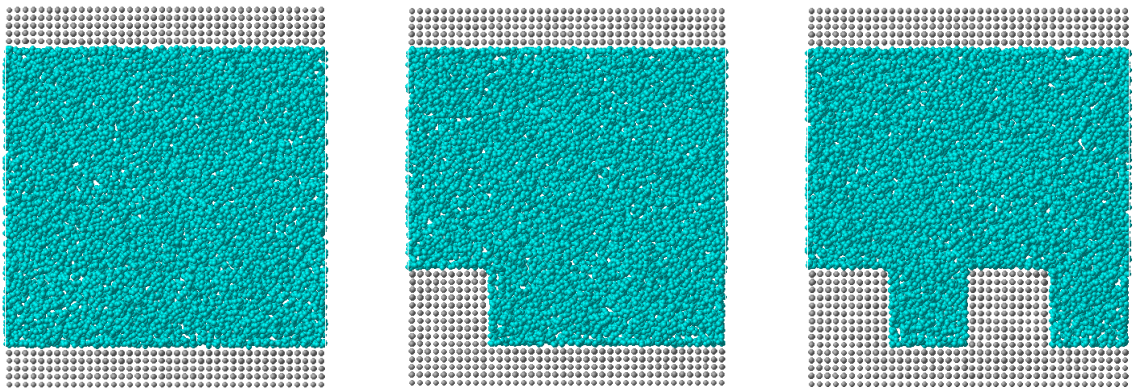

     **(a) Flat.**           **(b) Nano1.**           **(c) Nano2.**

**Figure 5. Snapshots of simulation system in the case of water state.**

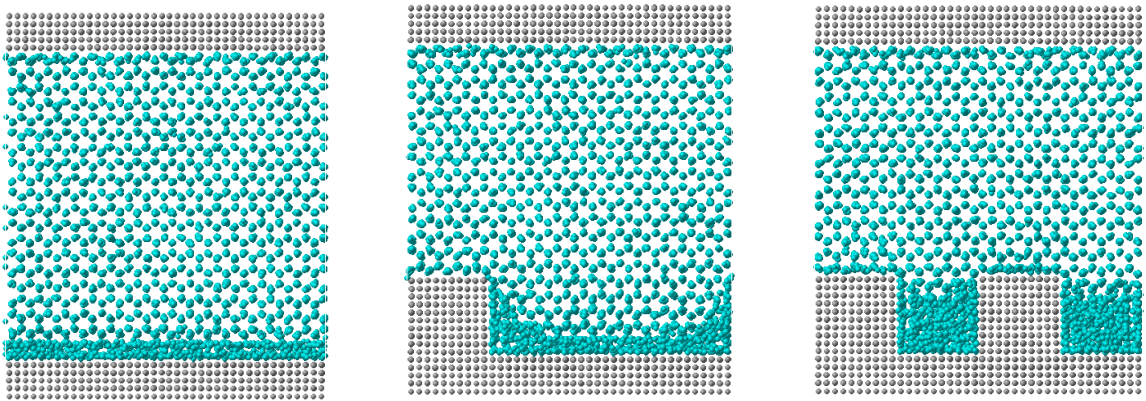

     **(a) Flat.**           **(b) Nano1.**           **(c) Nano2.**

**Figure 6. Snapshots of simulation system in the case of ice state.**

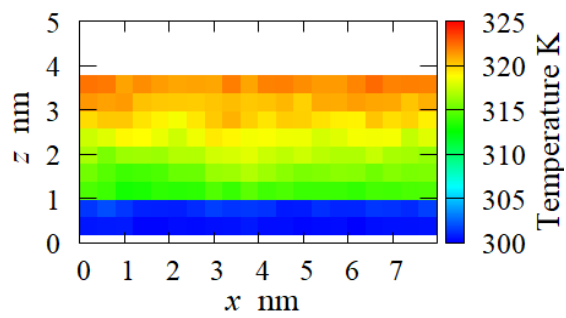

(a) Flat

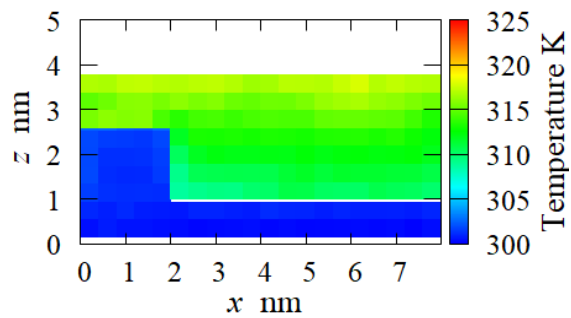

(b) Nano1

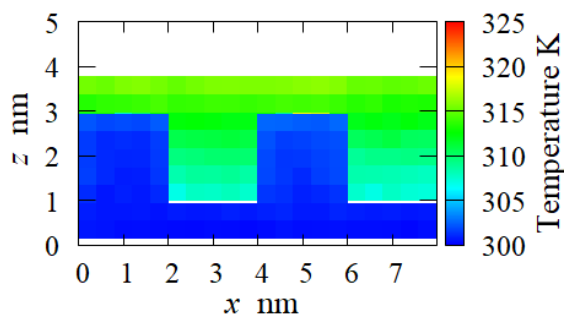

(c) Nano2

**Figure 7. Two-dimensional temperature distribution in water state.**

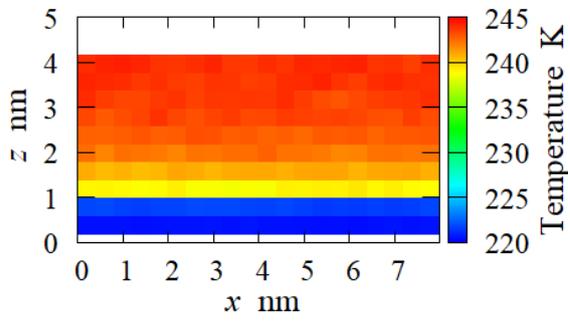

(a) Flat

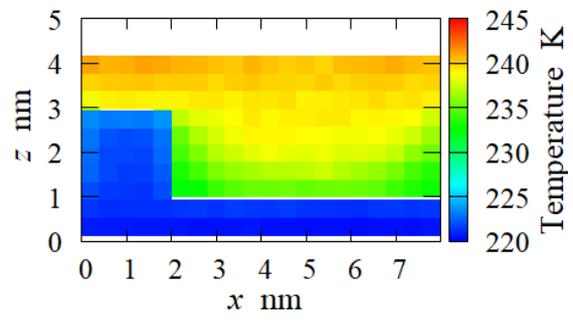

(b) Nano1

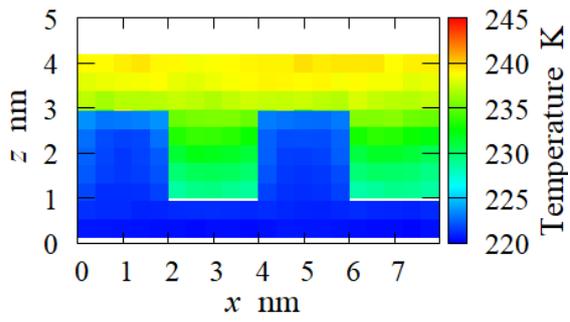

(c) Nano2

**Figure 8. Two-dimensional temperature distribution in ice state.**

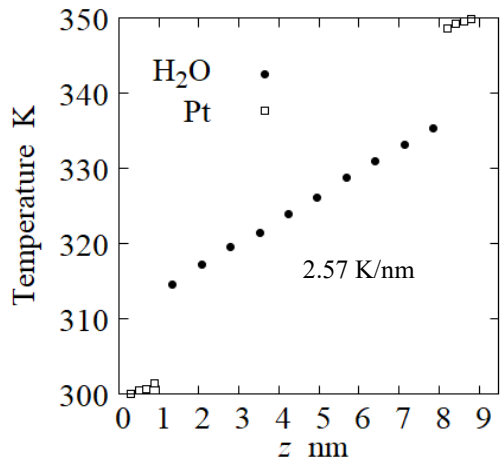
(a) Flat

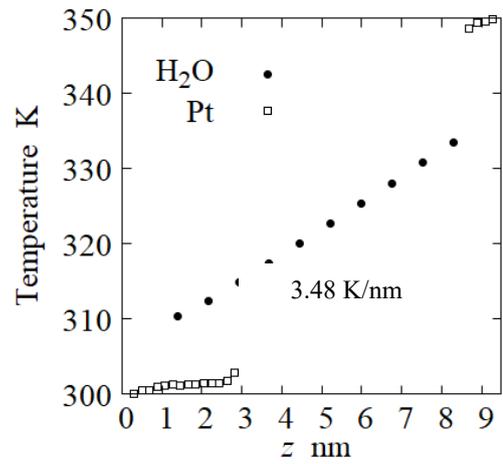
(b) Nano1

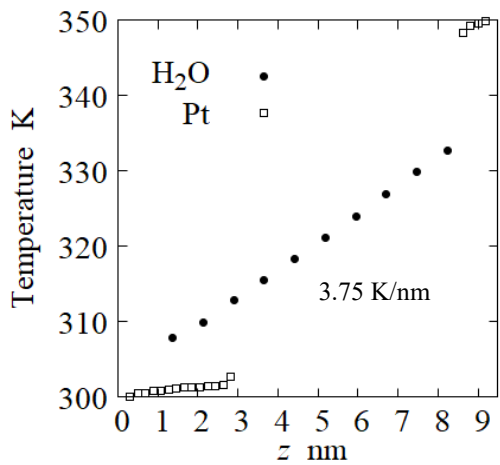
(c) Nano2

Figure 9. One-dimensional temperature distribution in water state.

(a)　　Flat

(b)　　Nano1

(c)　　Nano2

**Figure 10. One-dimensional temperature distribution in ice state.**

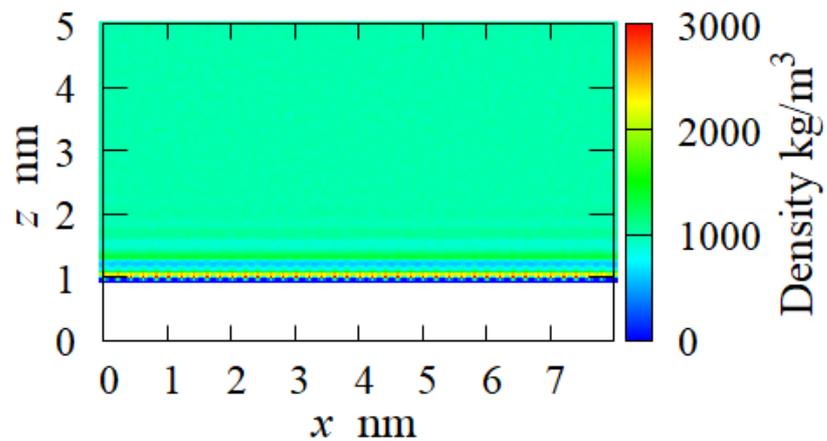
(a) Flat

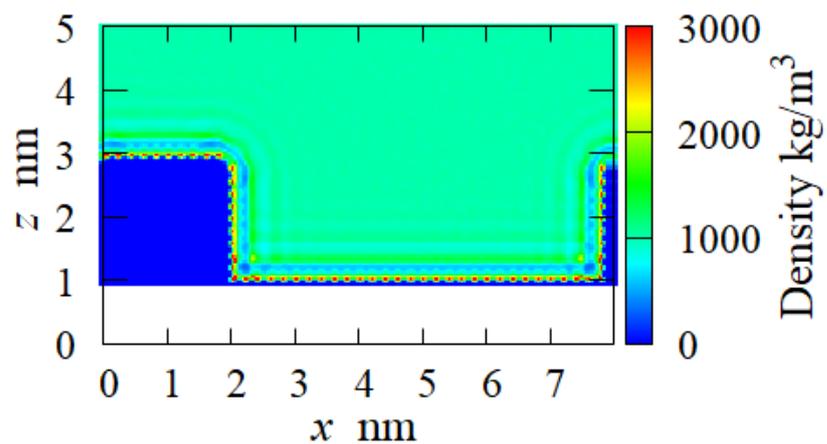
(b) Nano1

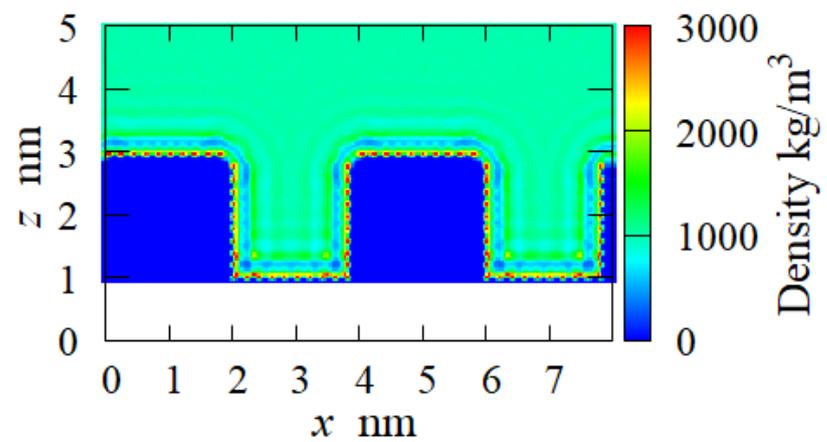
(c) Nano2

**Figure 11. Two-dimensional $H_2O$ molecules density distribution in water state.**

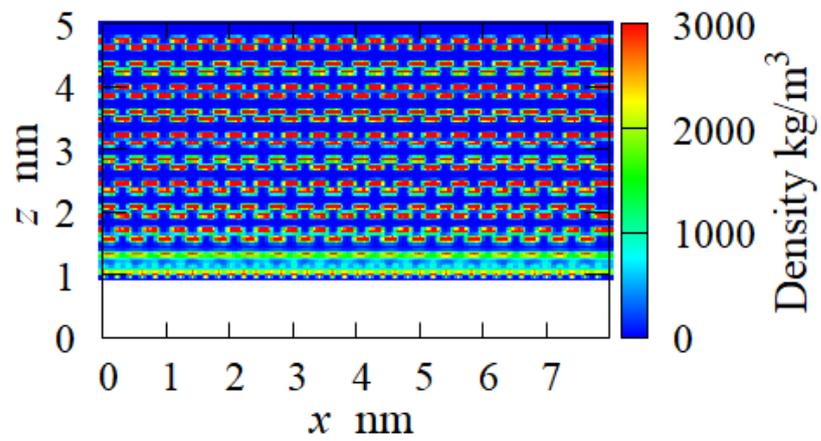

(a) Flat

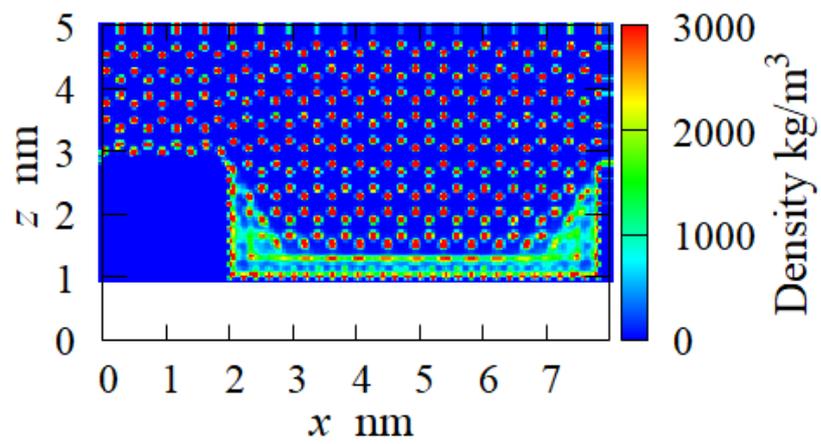

(b) Nano1

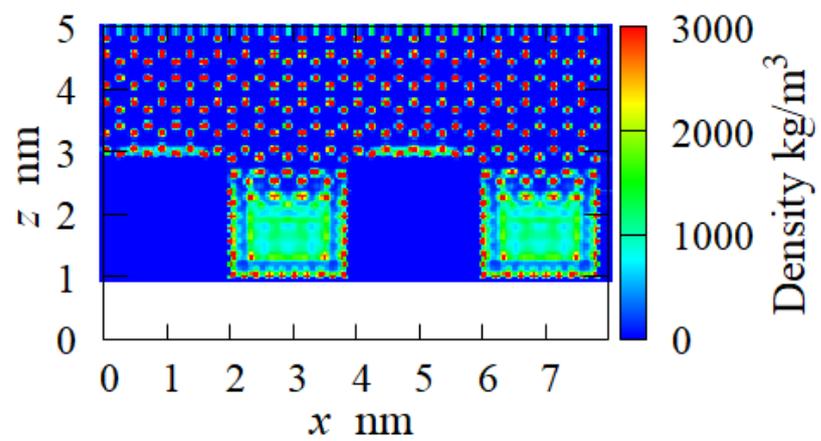

(c) Nano2

**Figure 12. Two-dimensional H$_2$O molecules density distribution in ice state.**

### 3.3 Thermal Resistance

**Figure 13** shows the thermal resistance of the interfacial region, $R_{th, region}$, in the water and ice states. The time-averaging was 5.0 ns. Based on them, we found that $R_{th, region}$ decreased as the nanostructure gap became narrower in both cases. It was because the narrower nanostructure gaps contributed to a larger volume of the intermolecular interaction between the $H_2O$ molecules and the Pt atoms, which induced higher heat flux through the interfacial region. In addition, we found that the $R_{th, region}$ was almost comparable in both the water and the ice cases.

**Figures 14** and **15** illustrate the local ITR, $R_{th, m}$, in the water and ice states. When the $H_2O$ molecules were in the water state, the local ITR of the bottom corner was relatively high in both cases of Nano1 and Nano2. On the other hand, the local ITR of the top corner was relatively low and even lower than that of the flat surface. When the $H_2O$ molecules were in the ice state in the case of Nano1, the local ITR of the bottom corner was relatively high, and that of the top corner was low. This tendency was similar to that in the case of the water state. However, the local ITR of the nanostructure top was significantly high. It was a different tendency from that of the water state. See **Figure 12** (b), and note that the crystal structure remained even on the nanostructure top. It induced a decrease in the local density and then an increase in the local ITR. In the case of Nano2, a location dependency of the local ITR was not significant. In between the nanostructures, the ice was in the amorphous form. Therefore, the tendency appeared in the cases of Nano1 and Nano2. Furthermore, in the case of Nano2, we did not find the tendency that the ITR of the nanostructure top increased. On the nanostructure top, the ice was not completely in crystal form. It indicated that the local density influenced the local ITR. The order of magnitude of the solid-liquid ITR was $10^{-9}$ - $10^{-7}$ $m^2K/W$, depending on the definitions of the thermal resistance, the surface wettability, and the fluid molecular models [e.g., 9-16, 21-24]. The order of magnitude of the ITR in the present study was consistent with the above-mentioned previous studies.

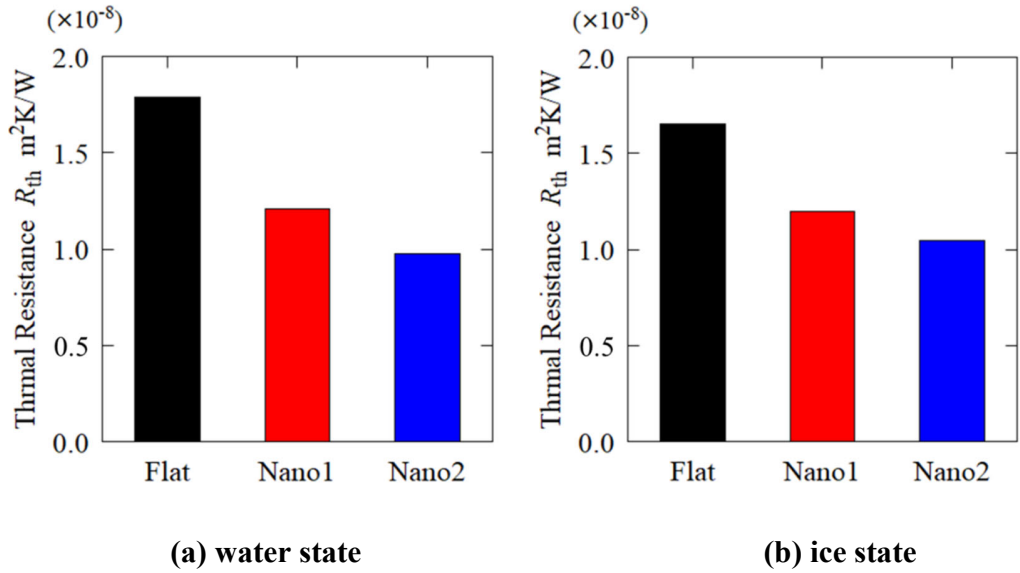

(a) water state     (b) ice state

Figure 13. Thermal resistance of interfacial region, $R_{th,\ region}$

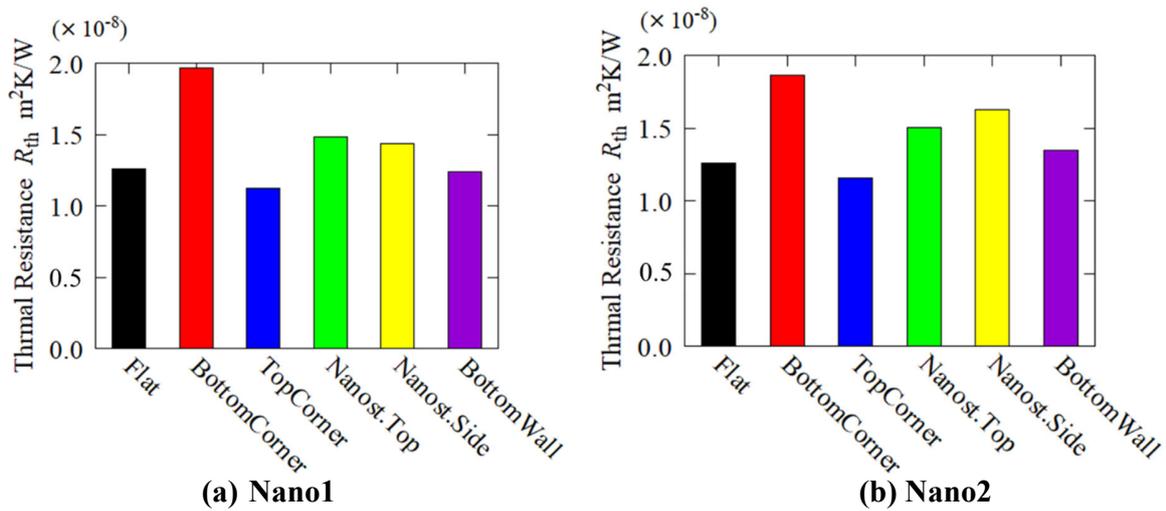

(a) Nano1     (b) Nano2

Figure 14. Local thermal resistance in water state.

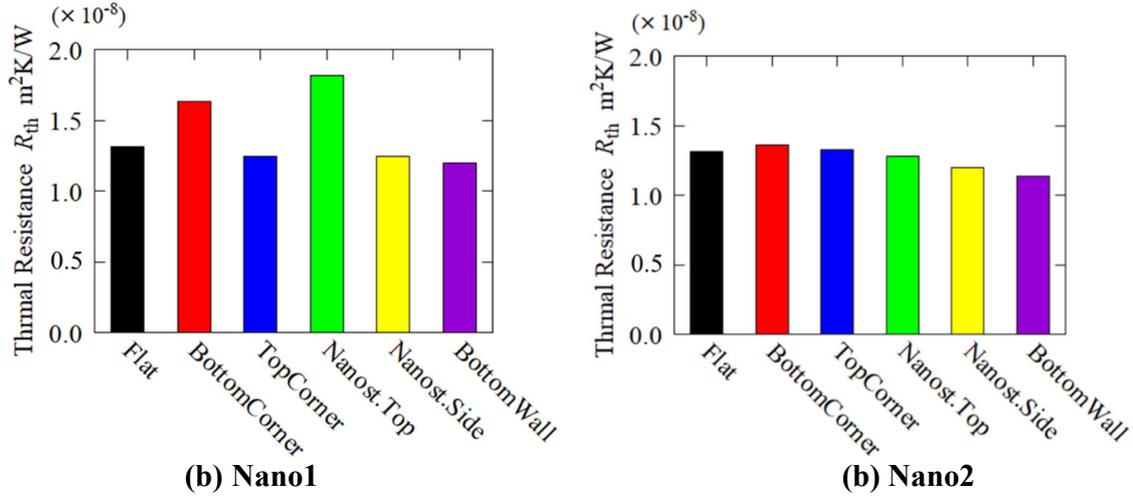

**(b) Nano1**  **(b) Nano2**

**Figure 15.** Local thermal resistance in ice state.

**3.4 Relation between Adsorption Layer and Interfacial Heat Flux**

The interfacial heaf flux was investigated in terms of a relationship with the adsorption layer of the $H_2O$ molecules. In the present study, the intermolecular potential for the $H_2O$-Pt pair was 12-6 Lennard-Jones potential multiplied with $\alpha$. At a boundary of the intermolecular distance of $r = 2^{1/6}\sigma$, the force between the pairs changed to be repulsive or attractive. If the intermolecular distance between the pairs were less than the boundary, the force on the pairs was repulsive. If it was greater, the force was attractive. With the present study parameters, the boundary of the intermolecular distance was 0.27725 nm. The local interfacial heat flux through the solid-liquid interface was evaluated from Eq. (8). The interfacial heat flux was discomposed into components of the attractive and the repulsive force contributions. **Figures 16** and **17** show the contributions of the attractive and repulsive forces to the interfacial heat flux in the water and ice states. From them, we found that the major of the interfacial heat flux was transported by the repulsive force of the intermolecular interaction regardless of the $H_2O$ molecules states.

For further understanding of the thermal energy transfer at the interface, a spatial decomposition of the interfacial heat flux was performed and shown in Figure 25. The computation system was the flat solid wall in contact with liquid $H_2O$ molecules. The flat solid wall was perpendicular to the z-direction. The spatial resolution was $2.00 \times 10^{-3}$ nm. The time-averaging was 2.0 ns. **Figure 18** shows the one-dimensional $H_2O$ molecules density distribution and interfacial heat flux decomposition as a function of the location of the $H_2O$ molecules around the first peak of the adsorption layer. The repulsive/attractive force boundary was located around the adsorption layer peak shown in **Figure 18**. Based on it, we found that the peak of the interfacial heat flux was not located on the position of the adsorption layer peak and that the $H_2O$ molecules, which were closer to the solid wall than the adsorption layer peak, transferred a major part of the heat flux. It was because the intermolecular distance got closer, and the intermolecular force became stronger. As a result, more amount of thermal energy was transported.

Then, we investigated the cases of the nanostructured surfaces. Figure 19 illustrates the schematic drawings of the spatial decomposition around the nanostructures for the density and interfacial heat flux distributions. The interrogation volume was in the L-shape in the *x-z* plane. Its thickness was $2.00 \times 10^{-3}$ nm. The local density and the interfacial heat flux were evaluated in each of the interrogation volumes. Here, *r* denotes the distance from the surface Pt atoms. The time-averaging was 2.0 ns. **Figure 20** shows that the one-dimensional $H_2O$ molecules density distribution and the interfacial heat flux decomposition as a function of the location of the $H_2O$ molecules around the first peak of the adsorption layer in the water state. From **Figure 20** (a), we found that the sum of the heat flux decomposition was 41% of that in the case of the flat solid wall. Note that the first peak of the adsorption layer was built up to be relatively high. It indicated that the density of the $H_2O$ molecules itself was not the dominant factor for the interfacial heat flux. At the nanostructure bottom corner, the intermolecular potential well was deeper than the others. As a result, the translation of the $H_2O$ molecules was restricted by the potential well. It might have influenced the local interfacial heat flux. From **Figure 20** (b), we found that the first peak of the adsorption layer became broader at the nanostructure top corner than that at the bottom corner and that the heat flux peak also became broader and higher. It meant that, around the top corner, the adsorption layer was formed more closely to the Pt atoms consisting of the solid walls, and then the proximity $H_2O$ molecules transferred more amount of the thermal energy through the interface. Therefore, the local ITR decreased in the water state. On the other hand, in the ice state, the adsorption layer was not formed along the nanostructure due to the crystallization, as shown in **Figure 12** (b). As a result, the local ITR did not significantly decrease.

**Figure 21** shows the one-dimensional $H_2O$ molecules density distribution and the interfacial heat flux decomposition as a function of the location of the $H_2O$ molecules around the first peak of the adsorption layer at the nanostructure top in the case of Nano1. From **Figure 12** (b), note that the $H_2O$ molecules in the ice state were in the crystal form just on the nanostructure top in the case of Nano1. In this case, the molecules' density peak became narrower and higher than that in the water state, as shown in **Figure 21** (b). It caused the proximity $H_2O$ molecules decreased in the ice state, and then the interfacial heat flux decreased by 58% with respect to that in the water state. Based on the mentioned-above we concluded that, regardless of the water or ice state, the greater amount of the thermal energy was transferred through the material interface by means of the intermolecular interaction when more the $H_2O$ molecules were located in the proximity area, which was closer to the Pt solid wall than the first adsorption layer peak.

In addition, the vibrational density of states (VODS) was locally evaluated for each of the interrogation volumes neighboring the water-solid and ice-solid interfaces, respectively. However, no significant relationship between the VDOS and the local ITR was found under the present numerical conditions. Because of this, the VDOS analyses were deliberately omitted from the present study.

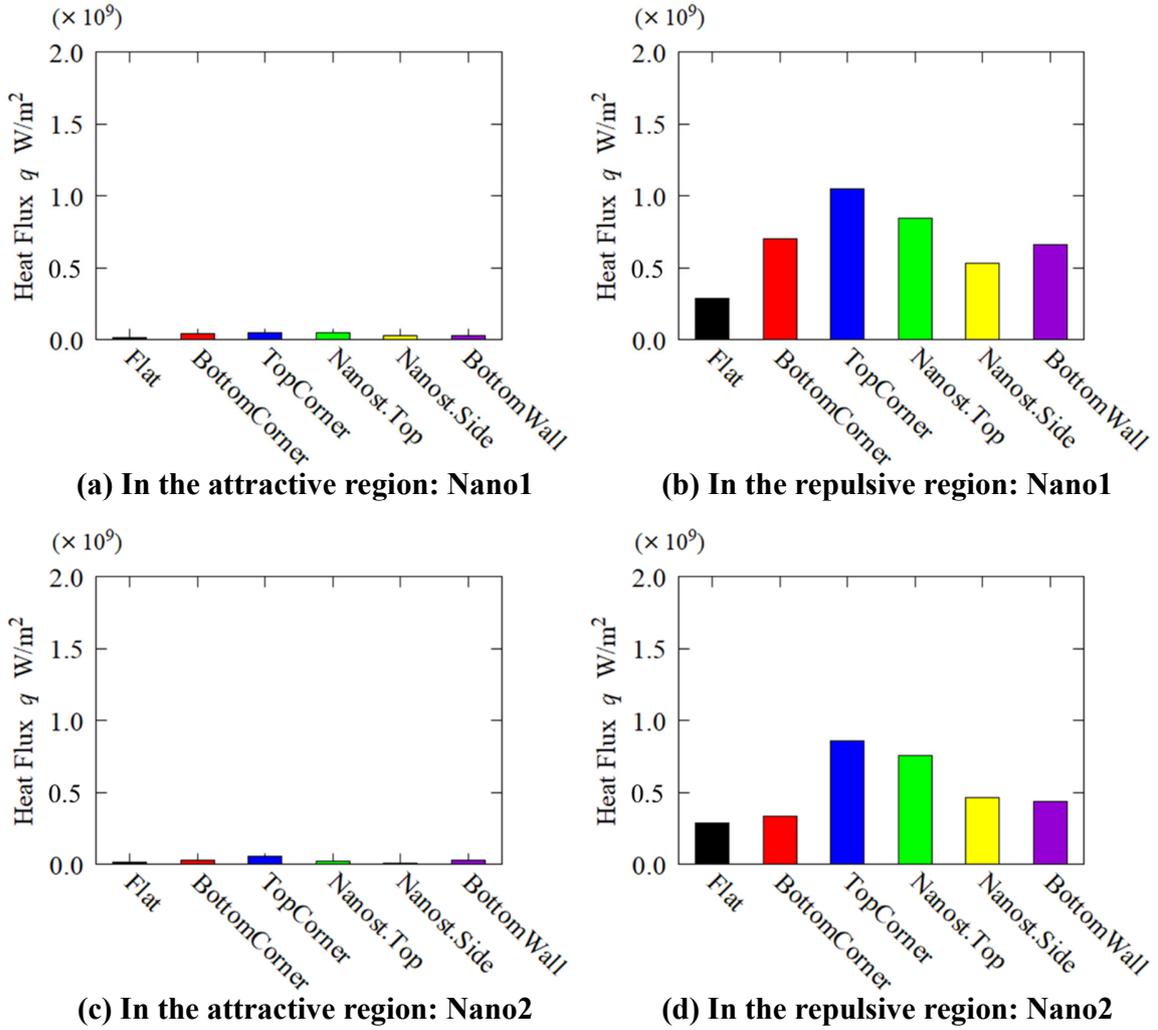

(a) In the attractive region: Nano1

(b) In the repulsive region: Nano1

(c) In the attractive region: Nano2

(d) In the repulsive region: Nano2

Figure 16. Contribution of attractive and repulsive forces to interfacial heat flux in water state.

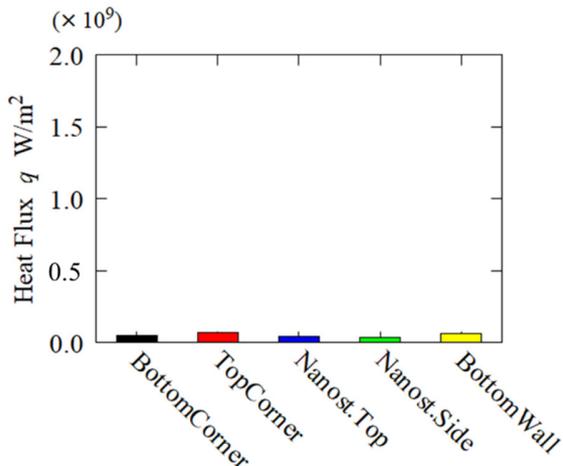 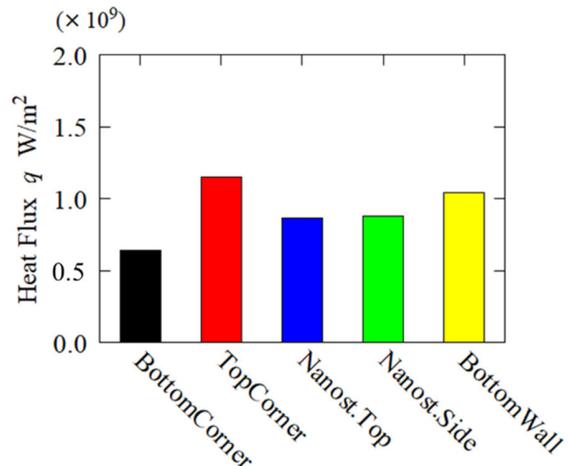

(a) In the attractive region: **Nano1**  (b) In the repulsive region: **Nano1**

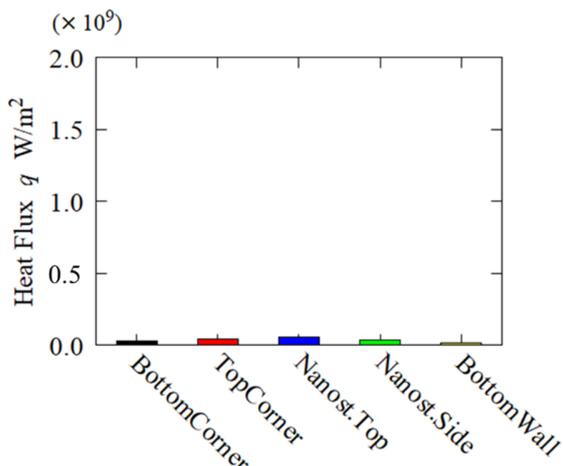 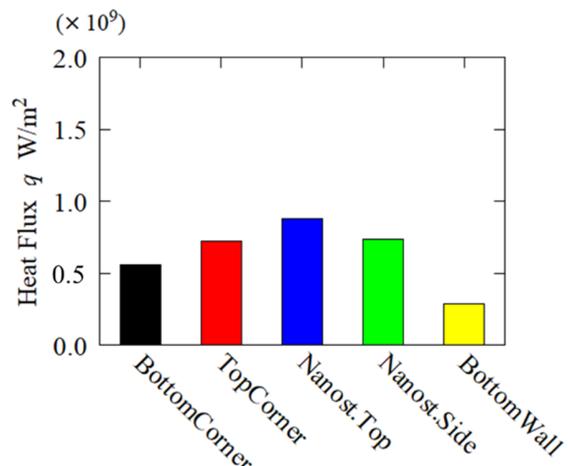

(c) In the attractive region: **Nano2**  (d) In the repulsive region: **Nano2**

**Figure 17.** Contribution of attractive and repulsive forces to interfacial heat flux in ice state.

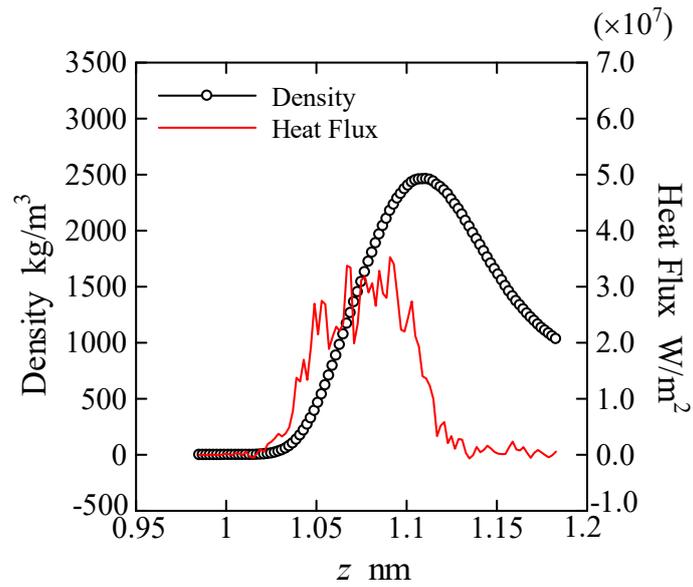

**Figure 18.** One-dimensional H$_2$O molecules density distribution and interfacial heat flux decomposition as function of location of H$_2$O molecules around first peak of adsorption layer in water state: Flat solid wall.

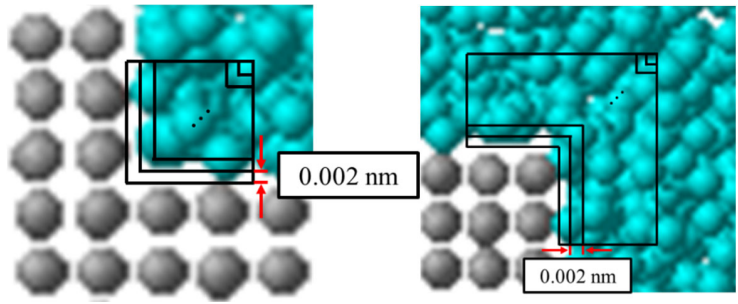

**Figure 19.** Spatial decomposition around nanostructures; Left: bottom corner; Right: top corner.

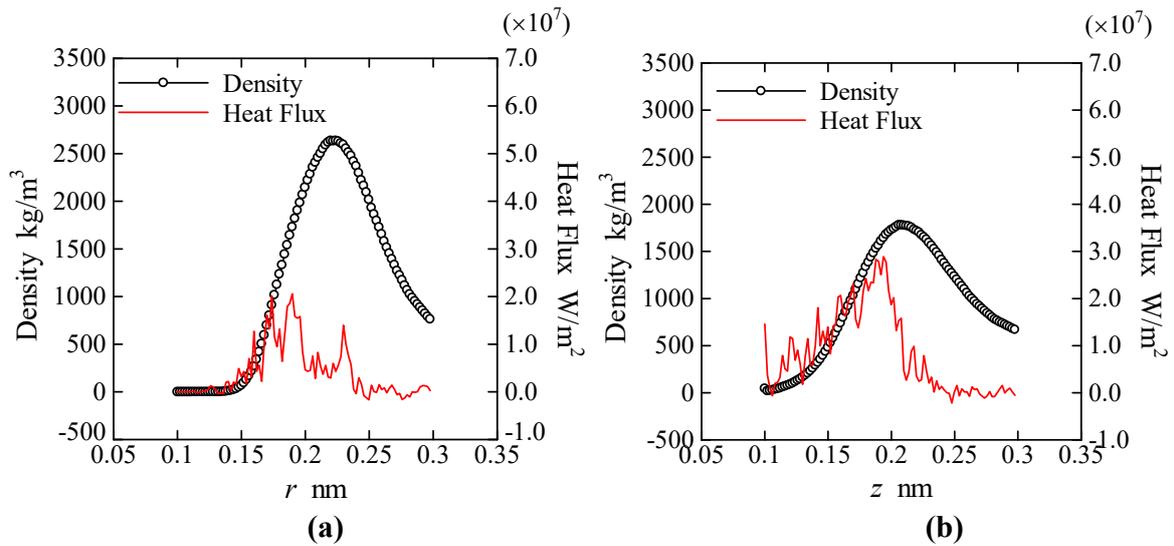

**Figure 20. One-dimensional H$_2$O molecules density distribution and interfacial heat flux decomposition as function of location of H$_2$O molecules around first peak of adsorption layer in water state; (a) At nanostructure bottom corner in the case of Nano1; (b) At nanostructure top corner in the case of Nano1.**

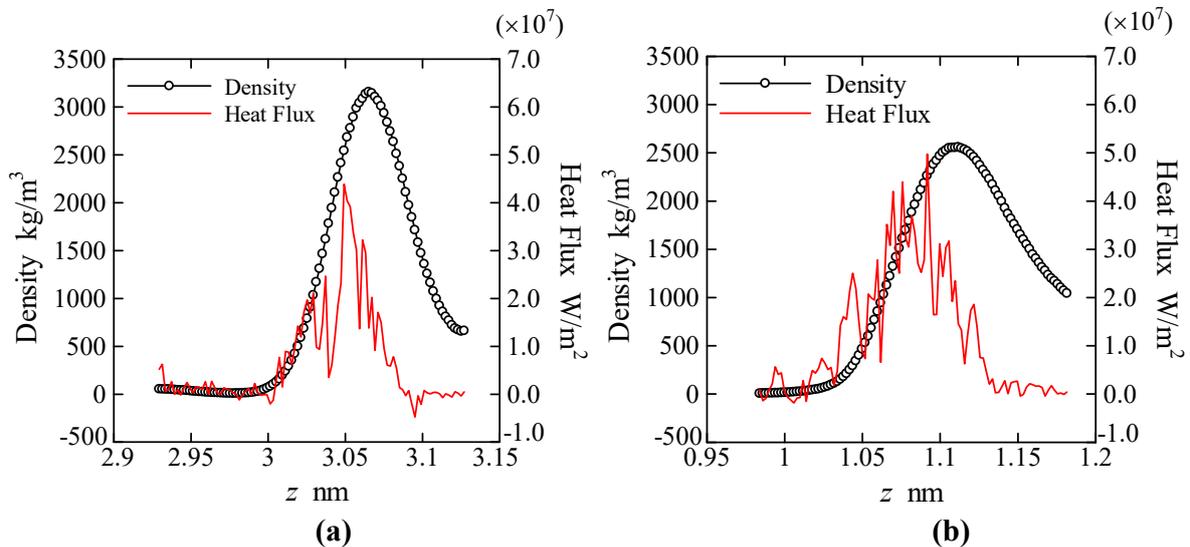

**Figure 21. One-dimensional H$_2$O molecules density distribution and interfacial heat flux decomposition as function of location of H$_2$O molecules around first peak of adsorption layer at nanostructure top in the case of Nano1; (a) ice state; (b) water state.**

4. CONCLUSIONS

In the present study, by means of the classical non-equilibrium molecular dynamics simulation, we investigate how and how much the nanotextured surface characteristics, such as surface wettability and geometry, influenced the interfacial thermal resistance between the solid wall and the water/ice. The mW $H_2O$ model and the LJ Pt model were employed. The intermolecular interaction between $H_2O$ and Pt was expressed by the LJ potential. Our findings are summarized as follows:

1. The ITR of the interfacial region was comparable in both the water and the ice states. As the nanostructure gaps became narrower, the ITR of the interfacial region decreased.

2. The local ITR was relatively high at the nanostructure bottom corner and low at the nanostructure top corner when the $H_2O$ molecules were in the water state. In the ice state, the local ITR was influenced by the ice structure neighboring the Pt solid wall. Especially when the $H_2O$ molecules were in the crystal form, the local ITR significantly increased.

3. A greater amount of the thermal energy was transferred through the material interface by means of the intermolecular interaction when more $H_2O$ molecules were located in the proximity area, which was closer to the Pt solid wall than the first adsorption layer peak. When the $H_2O$ molecules were in the crystal form on the solid wall, the proximity molecules decreased, and then the local ITR significantly increased.

**Appendix**

**A. One-Dimensional Density Distribution**

**Figures A1** and **A2** show one-dimensional density distributions of the water and the ice states in the *z*-direction. Note that, in the case of the water state, the molecules adsorption layers, where the density of the molecules surged, were formed on the flat solid wall (see **Figure A1** (a)). Hereafter, the density peak nearest the solid wall was called the first adsorption layer, and the second nearest peak was called the second adsorption layer. Notably, adsorption layers were found on the top of the nanostructures in the cases of Nano1 and Nano2 (see **Figure A1** (b) and (c)). The first adsorption layers formed on the top of the nanostructures were located at approximately *z* = 3.0 nm. The density peak became higher as the number of nanostructures increased. It was because, as the number of the Pt atom consisting of the nanostructures was increased, more $H_2O$ molecules were attracted. In between the nanostructures, some periodic density surges were found. We speculated that they mainly corresponded to the adsorption layers on the sidewalls of the nanostructures.

In the case of the ice state, the bulk regions were in the crystal form of the six-membered ring structure (see **Figure 12**). That typical feature was found in the present study (e.g., **Figure A2** (a)). On the bottom surface, the ice was in an amorphous form. Therefore, the first and second peaks were different from the density peaks in the bulk regions. In the case of Nano2, the ice in between the nanostructures was entirely in the amorphous form. Because of this, the density distribution significantly changed in between the nanostructures (see **Figure A2** (c)). However, the density distribution between the nanostructures in the case of the ice state had a different tendency from that in the water state (**Figure A1** (c)). It showed that the amorphous ice was different from the water confined between the nanostructures.

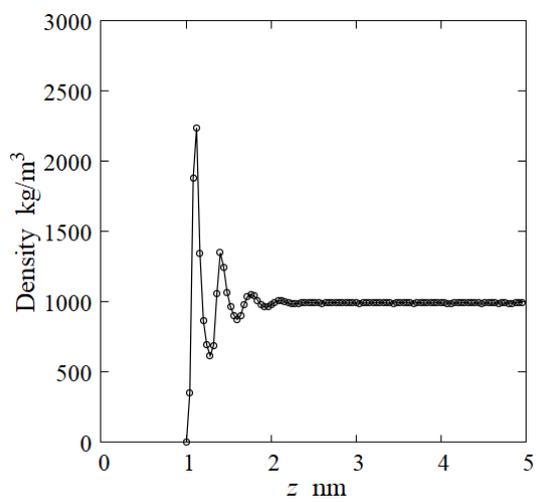
(a) Flat

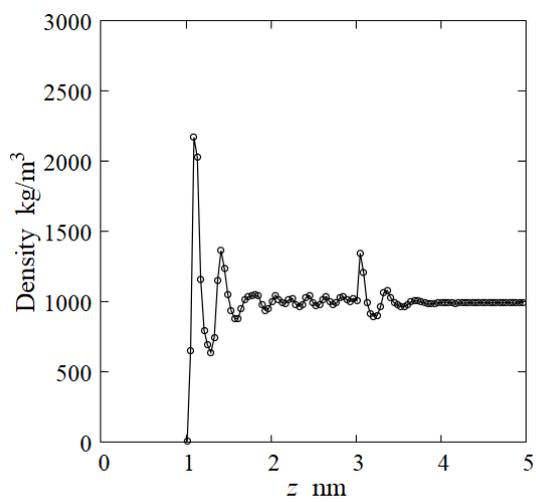
(b) Nano1

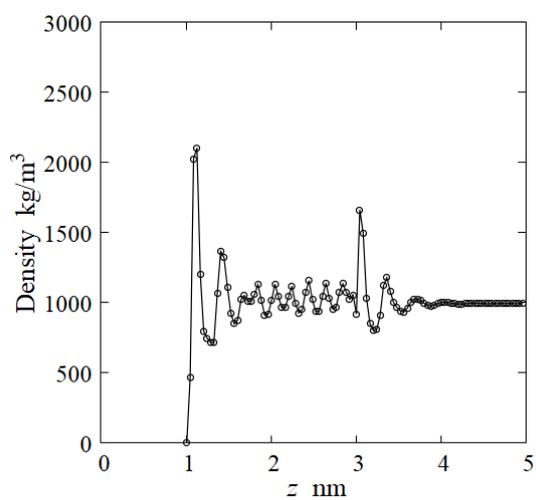
(c) Nano2

**Figure A1. One-dimensional $H_2O$ molecules density distribution in water state.**

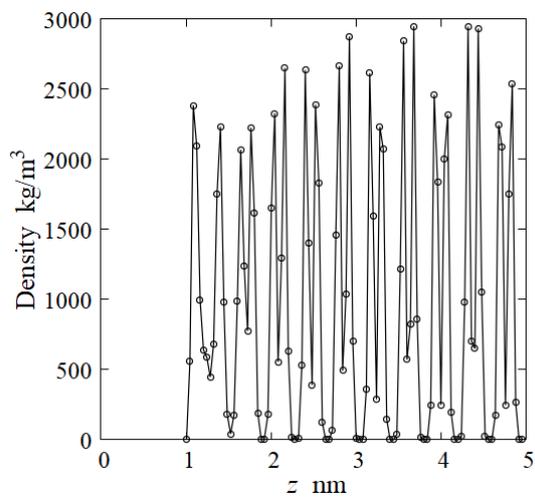
(a) Flat

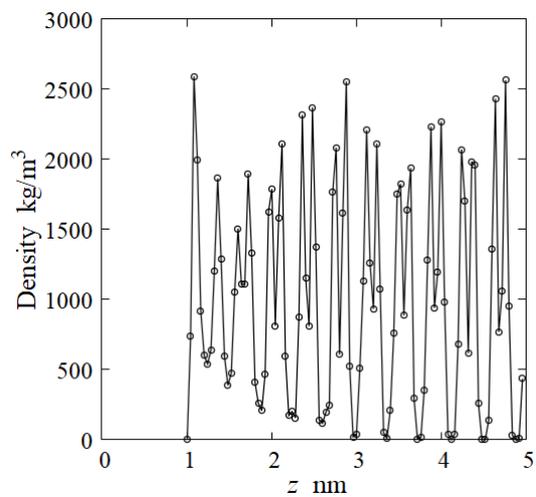
(b) Nano1

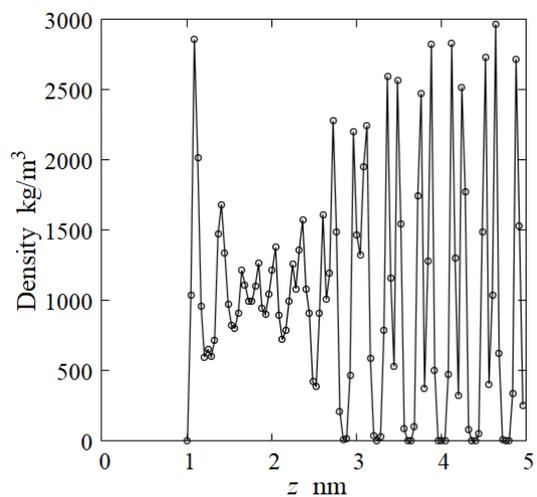
(c) Nano2

**Figure A2. One-dimensional $H_2O$ molecules density distribution in ice state.**

## B. Temperature Difference and Heat Flux of Thermal Resistance of Interfacial Region

**Figures B1** and **B2** show the temperature difference and the heat flux of the interfacial region in the water and ice states.

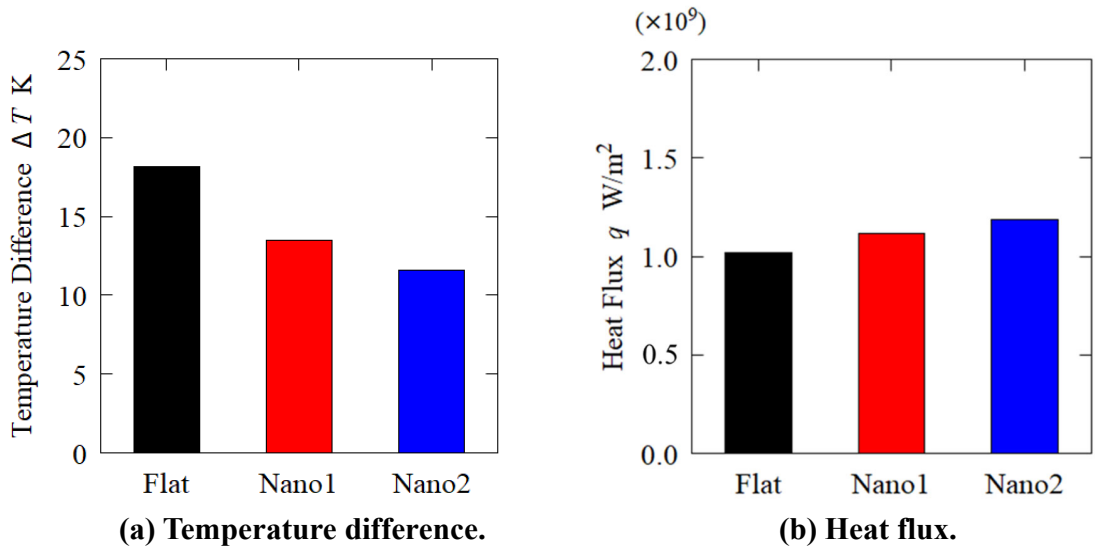

(a) Temperature difference.  (b) Heat flux.

**Figure B1. Temperature difference and heat flux of interfacial region in water state.**

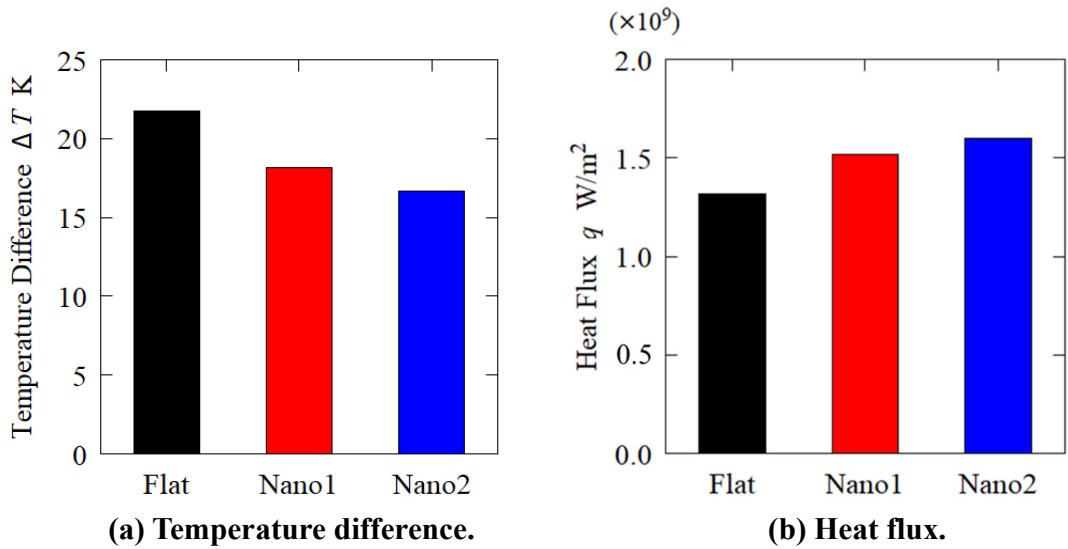

(a) Temperature difference.  (b) Heat flux.

**Figure B2. Temperature difference and heat flux of interfacial region in ice state.**

## C. Relation between Thermal Resistance and Density

The following two-dimensional density distributions were evaluated to investigate the relation between thermal resistance and density. The spatial resolution was 0.40 nm in a square in the *x-z* plane, which was identical to the interrogation volume for the local ITR. **Figures C1** and **C2** illustrate the two-dimensional density distributions in the water and ice states. We found that the local density was relatively low in the bottom corner, where the local ITR was relatively high, and that the local density was high in the nanostructure top corner, where the local ITR was low. In the case of the ice state, the local density was low in the bottom corner and the nanostructure top, where the local ITR was high. **Figure C3** shows the relation between the local ITR and the local density. Based on it, we found that a weak negative correlation between the local ITR and the local density, regardless of the states of the fluid molecules.

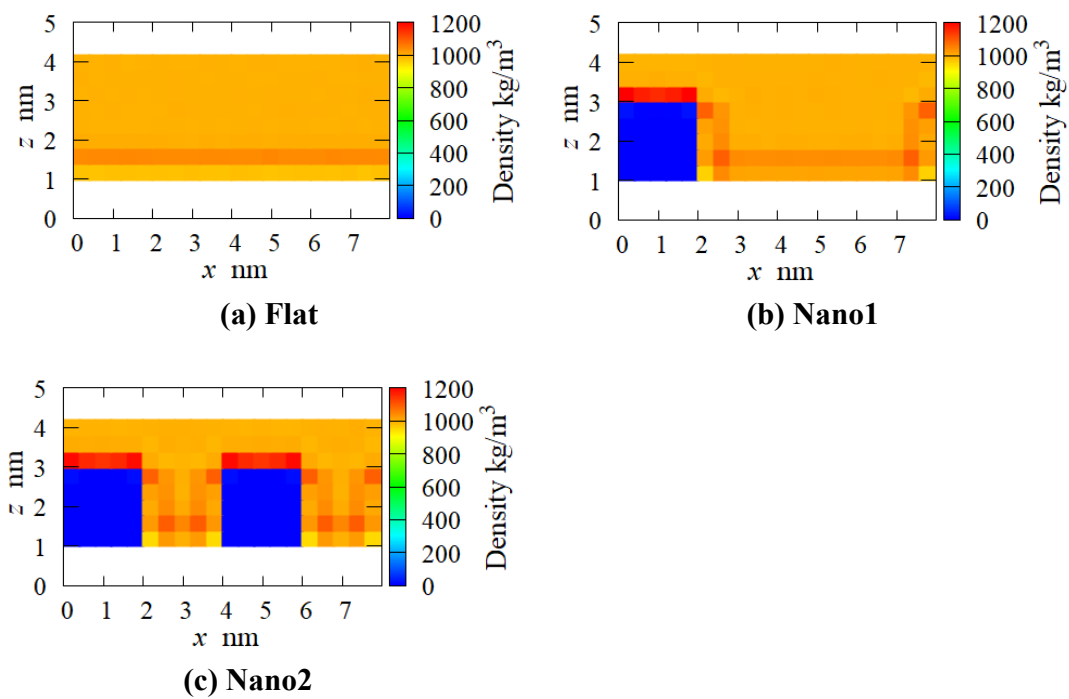

(a) Flat

(b) Nano1

(c) Nano2

**Figure C1. Density distribution in water state: spatial resolution of 0.4 nm.**

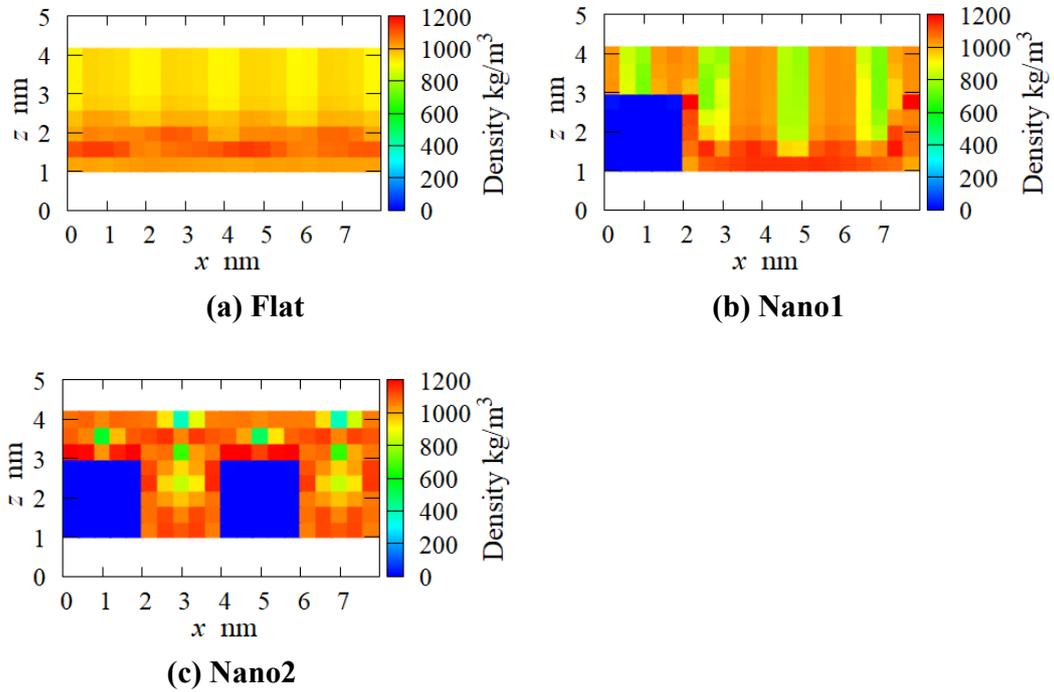

**Figure C2. Density distribution in ice state: spatial resolution of 0.4 nm.**

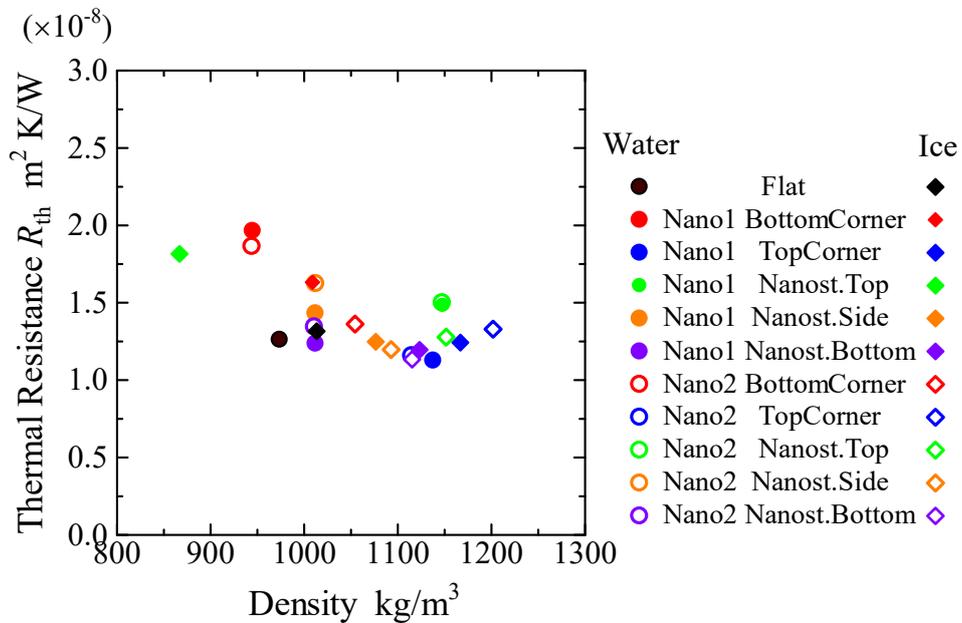

**Figure C3. Relation between local ITR and local density.**

AUTHOR INFORMATION

Corresponding Author

Yoshitaka Ueki – *Department of Mechanical Engineering, Osaka University, Suita, Osaka 565-0871, Japan; E-mail: ueki@mech.eng.osaka-u.ac.jp*

ACKNOWLEDGMENT

This work was supported by JSPS KAKENHI Grant Number 18H01382.